\documentstyle[12pt,epsfig]{article}
\begin{document}
 \baselineskip=4.8mm
\hskip 9.5cm DTP-95/86

\hskip 9.cm September, 1995

\vskip .8cm

{\centerline{\bf HADRON '95 --- SUMMARY~: PART I} }
\vskip 9mm

{\centerline{\bf M.R. PENNINGTON}}
\vskip 1mm

\baselineskip=4.5mm
\centerline{Centre for Particle Theory,
University of Durham,}

\centerline{Durham DH1 3LE, U.K.}
\vskip 6mm
\centerline{ABSTRACT}
\vskip 2mm
\baselineskip=4.5mm
\parskip=0mm
{\leftskip 1.4cm\rightskip 1.4cm{ Part I of this summary is
concerned with selected results on natural parity states presented
at this conference, in particular the glueball candidates the
$f_0(1510)$ and $\xi(2230)$. Unnatural parity and exotic states are discussed
 by Suh-Urk Chung$~^1$.}\par}
\vskip 3mm

\baselineskip=7mm
\parskip=2mm
\noindent {\bf 1. Introduction}
\vskip 2mm
\noindent The Summary of {\it Hadron'95} is the same as the summary of all
 previous Hadron conferences.
It is that QCD is the theory of the strong interactions.  Though we know
this to be true,
it is really only in very simple processes, where we have
hard scattering, that
we can compute with any degree of reliability what QCD has to say.  There,
 where
we have short distance interactions, we can use perturbation theory,
 make predictions,
compare them with experiment and find that they agree to a $K$-factor or two.
Such perturbative ideas govern the short-distance part of the
inter-quark potential, where simple one gluon exchange dominates,
 thanks to asymptotic freedom, Fig.~1.
 \begin{figure}
\vspace*{16cm}
\caption[Fig.~1]{The inter-quark potential as a function of separation $r$.
 It is controlled by one
gluon exchange at short distances and multigluon exchange at long range.}
\end{figure}
\noindent However, the bulk of hadronic phenomena are governed by the
 long-distance regime
controlled by confinement. There a whole mesh of gluons are exchanged to
 produce the force between two
quarks, Fig.~1, and we really do not know how to calculate what is going on.
 It is this
region that determines the  spectrum of light hadrons and
indeed all confinement physics.
It is from experiment that we primarily learn about this regime.
There hadron spectroscopy
is the natural guide. It  governs not just low energy hadron and nuclear
processes,
but even high energy scattering. Though the total cross-section
for $e^+e^-$ annihilation
may be treated perturbatively and even the cross-section for jet production,
as soon as we ask the question \lq\lq what is the probability of finding
some specific hadron, like a pion, in a jet", we must confront confinement
and GeV scale physics. The place to learn about this is from the spectrum of
light hadrons and in particular   in the meson sector, as this is where
considerable progress has been made.

  We begin our discussion of light mesons with the simple
quark model picture
 with three flavours. Here a quark and antiquark are assumed bound into states
whose quantum numbers are determined by the spin, $S_{q{\overline q}}$,
of the $q{\overline q}$ system and the relative orbital angular momentum,
$L_{q{\overline q}}$, of the quark and antiquark. This leads to the
 familiar multiplet structure, so readily seen
for pseudoscalars, vectors and tensor mesons. Moreover, the mass of an
$L_{q{\overline q}}=0$ meson like the $\rho$, made of two constituent quarks,
is just two-thirds of the mass of the nucleon made of three such quarks.
This picture has proved a valuable aid to our understanding.

  However, we now have QCD.  This seems to complicate matters enormously.
A constituent quark, we learn, is really a current quark surrounded
 by a cloud of gluons
and a sea of $q{\overline q}$ pairs. The success of the  naive
 quark picture means
  that this cloud of gluons is just the same in a $\rho$ meson as in a proton.
  This gluonic component is in some way universal. The belief
  that colour is confined and consequently hadrons are colour singlets
  gives us an understanding of why the mesons we see are
   made of a quark and an antiquark and baryons are made of three quarks.
   However, QCD leads us to
   expect a far richer spectrum of colour singlet states  with mesons
   made of more quarks, such as $qq{\overline {qq}}$, or hybrid mixtures of
   quarks and gluons, such as $q{\overline q}g$, and
    even states with no quarks  at all ---
   glueballs, such as $gg$. Indeed, QCD demands that such states must exist.
  Thus, the main thrust of experimental studies of the hadron
   spectrum has for the last twenty years been the search for unambiguous
   evidence for states beyond the quark model.
\vskip 5mm
\noindent {\bf 2. Scalar mesons}
\vskip 2mm
\begin{figure}
\vspace*{17cm}
\caption[Fig.~2]{The mass spectrum of the $I=0$, $I=1/2$ and $I=1$ $0^{++}$
mesons.  Those with the black dots have been discussed at this conference.}
\end{figure}

 \begin{figure}[th]
\vspace*{7cm}
\caption[Fig.~3]{Unitarity requires a resonance, $R$, that decays
 to $\pi\pi$, for example,
has to couple in the same way to this final state whether produced in $\pi\pi$
scattering or centrally in $pp\to pp(\pi\pi)$ via a double Pomeron mechanism.}
\end{figure}

\noindent  The scalar meson sector is the one that has received
most attention at this
conference.  In Fig.~2 is shown the mass spectrum of $I=0$, $I=1/2$  and $I=1$
$0^{++}$ states. Those with the black dots
alongside have been discussed at this conference$~^{2-11}$.
The first thing to decide is how many of these are real
and how many are distinct. The candidates for an $f_0(500)$ and $f_0(750)$
(often called $\sigma $'s) must, I believe, be unphysical.
 It is worth spending a minute on this. Hadron states correspond
  to poles of the $S$-matrix on the nearby unphysical sheet.
 The first remark is that they  need not have anything  to do with
 poles of the $K$-matrix.  The $K$-matrix is just a convenience and not a
 physical quantity.  The fact that probability must be
  conserved in any process means
 that the $S-$matrix must be unitary, i.e. $S^{\dag}S\,=\,1$.  Unitarity
 demands that resonance poles occuring in one channel must appear
 in all other channels with
 the same sets of quantum numbers.  This universality means that a resonance
 that appears
 in central dipion production in $pp$ scattering must also appear in
 $\pi\pi\to\pi\pi$, Fig.~3. It just cannot avoid this.  Thus claims of a
 narrow $\sigma(500)$
 in the GAMS results$~^7$ cannot be correct as no such state is seen in
 $\pi\pi$ scattering.  Unitarity demands a universality that requires
 central production, for instance, to be analysed in a way consistent
 with other
 information on the same channels and not in isolation.
 I believe we can therefore
 discount the $f_0(500)$, and in a similar way the $f_0(750)$ $~^8$, which
 is inconsistent with $\pi^0\pi^0$ production in the
  BNL E852 experiment$~^{12}$.
 The broad scalar state in Fig.~2, denoted by 800-1300,
 the $f_0(\epsilon(1000))$, is what PDG'94$~^{13}$ calls
 the $f_0(1300)$.

  Let us turn to the other isoscalar states.
As shown in the talks of Stefan Spanier$~^2$ and Stefan Resag$~^3$, the
 Crystal Barrel collaboration have studied ${\overline p}p$
 annihilation at rest into
 $(\pi^0\pi^0)\pi^0$, $(\eta\eta)\pi^0$ and $(4\pi^0)\pi^0$.  Analyses of their
 beautiful Dalitz plots, assuming that the  ${\overline p}p$
 annihilate purely in a $^1 S_0$ state, requires an $f_0(1510)$ with a width of
 $150 \pm 30$ MeV, in addition to an $f_0(1370)$,
 with a width of $\sim 350$ MeV and  a much
 larger branching ratio to $4\pi$ than to $2\pi$.
 With 170,000 events in the $3\pi^0$ channel and a similar number
 with $\eta\eta\pi^0$, $S-$wave signals at 1370 and 1510 MeV
 look very impressive. In the $5\pi^0$ channel, the $f_0(1510)$ signal
 is  less dramatic giving an improvement in the maximum likelihood for both
 mass and angular distributions over and above
 the broad $f_0(1370)$.

   A major issue is whether
 the $f_0(1370)$ is really distinct from the equally broad $f_0(1300)$ or not.
 The answer to this question depends on the particular analysis of the
 Crystal Barrel results$~^{14}$.  There are those who say that
 a single broad scalar
 stretching from 700 MeV, where it is totally elastic, up to 1400 MeV,
 where it has become highly inelastic because of the opening of the $\eta\eta$
 and appreciable $4\pi$ channels, is just as consistent with data
 as a specifically inelastic $f_0(1370)$ on top of a slowly varying but large
 background and those who are adamant that a single broad state
 does not describe the same
 results$~^{14}$. Let us leave this to be resolved.

   Whether the $f_0(1510)$ of Crystal Barrel is the same as the
 $f_0(1525)$ found by LASS$~^{15}$ and the $f_0(1590)$ of
  GAMS$~^{16}$ has also been much discussed.
 Here there is at least a conscensus that they may well be the same.
 The evidence for an $f_0(1525)$ under the well-known $f_2(1525)$,
 with essentially the same mass and width as the leading $D$-wave looked
 highly speculative even in the results of such a high statistics experiment
 as LASS$~^{15}$. But now that Crystal Barrel has definitely seen a scalar
  in this mass
 region, it is time to go back and perform a common analysis of this
 mass region.

   Such a combined treatment is the aim of three
 analyses presented here by Anisovich et al.$~^4$, Bugg et al.$~^5$,
 and Anisovich et al.$~^6$,
 in which Sarantsev was a collaborator in each.  The aim was
 to treat the classic  data on peripheral dipion production
 in $\pi^-p\to\pi^-\pi^+ n$, $\to\pi^-\pi^-p$ from the
 CERN-Munich group$~^{17}$, on $\pi^-p\to\pi^0\pi^0 n$ from GAMS$~^{18}$,
 the Mark III results$~^{19}$ on $J/\psi\to\gamma\pi^+\pi^-\pi^+\pi^-$
 together with the Crystal Barrel Dalitz plots on
 ${\overline p}p\to\pi^0\pi^0\pi^0$, $\to \eta\eta\pi^0$$~^{20}$ and
 $\eta\pi^0\pi^0$$~^{21}$.
 These simultaneous analyses conclude that there are four $0^{++}$ isoscalars~:
 the $f_0(980)$, $f_0(\epsilon(1000))$, $f_0(1370)$ and the $f_0(1510)$.
 An important part of this is the revised opinion of the quantum numbers of the
 $4\pi$ signal in $J/\psi$ radiative decays.  This is an illustration
  of the {\it metamorphosis}
 that Achasov$~^{22}$ talked about and highlights the uncertainty
  in quantum number determination
 in complex final states.  In these analyses the $\pi\pi\to\pi\pi$ $S$-wave
 cross-section not only
 has the well-known sharp drop  because of the $f_0(980)$ but an
 analogous one for
 the $f_0(1510)$. A surprising omission in this treatment
 is information on   $K{\overline K}$ final states,
 as a major source of inelasticity
 in the $\pi\pi$ and $\eta\pi$ channels. Forthcoming results from
Crystal Barrel may help here too.
\vskip 5mm
\noindent {\bf 3. Which are the $q{\overline q}$ scalars~?}
\vskip 2mm
\baselineskip=7.1mm
\noindent Returning to Fig.~2, the next question to ask is   how many
$q{\overline q}$ scalar multiplets are there~?  Indeed,
 which states belong to the lowest
lying nonet~?  The fact that below 1800 MeV there is only one $K_0^*$,
the $K^*_0(1430)$ found so cleanly by LASS$~^{23}$,
unambiguously points to only
one $q{\overline q}$ multiplet in this mass region
despite the two $a_0$'s and very
many $f_0$'s. To help decide which are $q{\overline q}$ states, let us consider
a case study.
\begin{figure}[th]
\vspace*{8.5cm}
\caption[Fig.~4]{$1^{--}$ ideally mixed $q{\overline q}$ nonet and the
corresponding meson states.}
\end{figure}

  We start from the well-known nine lightest vector mesons with the spectrum
shown in Fig.~4. There is a one-to-one correspondence
between the observed mesons
and the underlying ideally-mixed quark model nonet. This is the success of the
quark model. Let us see why.
One begins with the bare quark state propagators with denominators
$(m_0^2\,-\,s)$. Rather like a $K-$matrix element these bare propagators
have poles on the real axis at $s\equiv E^2 = m_0^2$.
\begin{figure}
\begin{center}
\mbox{\epsfig{file=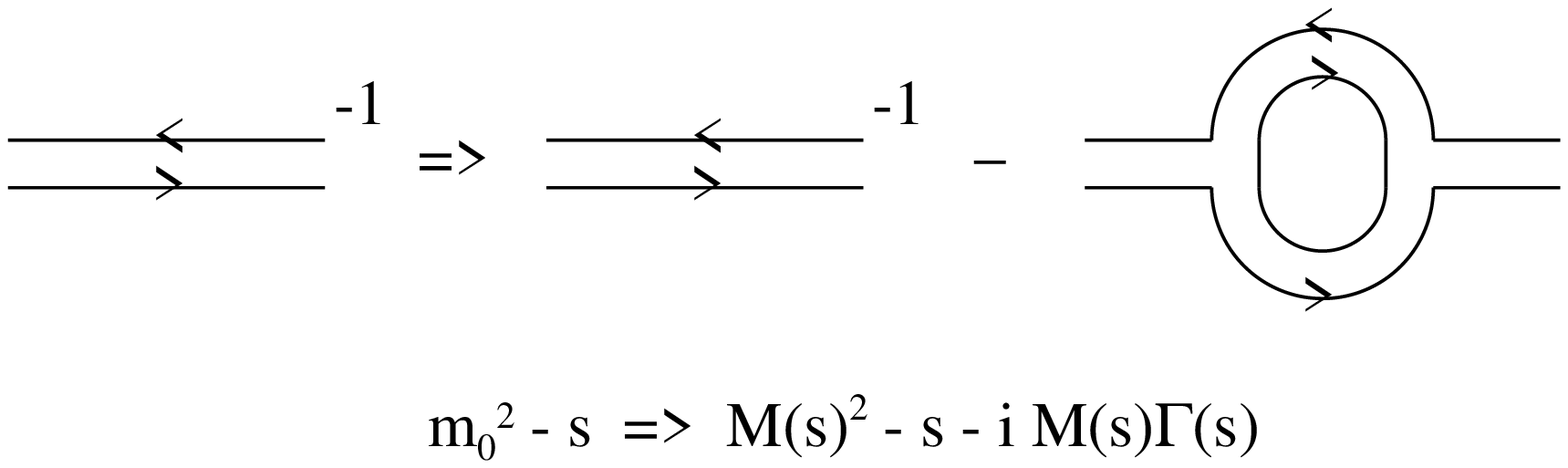,height=4cm}}
\end{center}
\caption[Fig.~5]{How the inverse propagator for a bare $q{\overline q}$ state
changes when its couplings to hadrons are included.}
\end{figure}
 Now we switch on interactions, Fig.~5.
This allows the $q{\overline q}$ states to couple to hadronic
final states (which is what
accounts for their decays)
and produces hadron loops in their propagators.  The effect of these
 on the inverse propagators
is to change the zeroth order result $(m_0^2\,-\,s)$ to
${\cal M}(s)^2\,-s\,-i{\cal M}(s) \Gamma(s)$.  The poles in the propagator
have moved into the complex plane as displayed in Fig.~6 and there are branch
 cuts at each hadronic threshold.  Now the $I=0$ states, for example,
  can couple to $3\pi$
 and $K{\overline K}$.
 \begin{figure}
\vspace*{6.cm}
\caption[Fig.~6]{The complex $s$-plane for the states of Fig.~5. The bare
propagator has a pole on the positive real axis.  The dressed propagator
has a pole below
the real axis and the plane has cuts generated by the hadronic
 intermediate states in
the loop of Fig.~5.}
\end{figure}

 The bare propagator for an $s{\overline s}$
  corresponds to a non-decaying $\phi$, i.e.
  $$ \vert \phi \rangle_0\;=\;\vert s{\overline s}\rangle\qquad .$$
Turning on interactions one finds that
the physical $\phi$ has a Fock space decomposition as
$$\vert \phi \rangle_1\;=\;\sqrt{1-\epsilon^2}\,
\vert s{\overline s}\rangle\;+\;
\epsilon_1 \vert K{\overline K} \rangle\,+\,\epsilon_2 \vert \rho\pi \rangle\,
+ ...\; ,$$
where $\epsilon^2 = \epsilon_1^2 + \epsilon_2^2 + ... \ll 1$.
Consequently, the physical $\phi$ is overwhelmingly an $s{\overline s}$
state, just like the bare one, and the switching on of interactions
produces a relatively small effect.  Then the simple quark picture works.
We observe hadrons, but we can nevertheless readily infer
their quark substructure.
A similar picture holds for tensor mesons.  Now let us turn to the scalars.
\begin{figure}[th]
\vspace*{11cm}
\caption[Fig.~7]{$0^{++}$ ideally mixed $q{\overline q}$ nonet with
 the corresponding meson
states in the dynamical scheme of Tornqvist$~^{9}$.}
\end{figure}

  As an illustration, we start from an ideal quark model nonet
as in Fig.~7
with the non-strange quark states around 1420 MeV and replacing the $u$ or $d$
with an $s$ quark adds 100 MeV to the mass.  Let us now turn on the coupling
to hadrons as in Fig.~5, but this time just to two pseudoscalars.
 This is the explicit calculation performed by Tornqvist$~^{9}$.
What makes the situation different from that for vectors and tensors is firstly
that
the coupling to $\pi\pi$, $\eta\eta$ etc. are larger and secondly
the thresholds for $0^{++}\,\to\,0^{-+}\,0^{-+}$ are $S-$wave.
So including just the two pseudoscalar decays, the outcome is dramatically
different from that outlined for the vectors. The physical states are
 shown in Fig.~7,
in which the $f_0(1200/1300)$ is broad and the
 $a_0(980)$ and $f_0(980)$ appear narrow because of their
 proximity to $K{\overline K}$ threshold. These first approximations to the
 physical states can similarly be decomposed into Fock space. Now,
 for example~:
 $$ \vert f_0(980) \rangle_1\;=\; \epsilon_1 \vert s{\overline s} \rangle\;+\;
 \sqrt{1-\epsilon^2} \vert K{\overline K} \rangle\;+\;...\quad .$$
 Once again $\epsilon_1 ^2 < \epsilon ^2 \ll 1$, but now the $f_0(980)$
 is largely a $K{\overline K}$ state and not an $s{\overline s}$.  This
 should
 not be confused with a $K{\overline K}$ molecule, as here the {\it seeds}
 are definitely quark model states, and the $f_0 (980)$ is not  bound
  by inter-hadron forces alone$~^{24}$.

  Tornqvist's valuable model calculation
 serves as a serious warning. It highlights
 how in the scalar sector no
 close relationship between the observed hadrons and the underlying
 quark states  is to be simply seen.  It is, however, important to
 recognise that Tornqvist's calculation is a model~:
 a particular unitarisation is used, only two pseudoscalar channels,
 $\pi\pi$, $K{\overline K}$, $\pi\eta$, ... , are included and
 there are only the  nine {\it seeds}  of a simple quark model multiplet.
 Questions abound~:
 what is the role of $4\pi$ channels, which for primitive states
 at 1500 MeV must be important~?
 What happens if one introduces a purely gluonic {\it seed} too~?
 This is clearly a territory awaiting further exploration.

\begin{figure}
\vspace*{17.5cm}
\caption[Fig.~8]{Comparison of the observed spectrum of $0^{++}$ mesons
with (I) the scheme supported by Tornqvist's calculation$~^{9}$ and (II)
the scheme of Amsler and Close$~^{10}$.}
\end{figure}

At this conference, we have also heard about the work of
Amsler and Close$~^{10}$ on the same $0^{++}$
sector.  The quark model {\it seeds}  are here nine $q{\overline q}$'s and
one $gg$.  The mixing of these  primitives  is calculated in
old-fashioned perturbation theory.  The outcome is that
the $f_0(1510)$ is the predominantly glue state.
 This calculation is non-relativistic and  not
at the level of sophistication of Tornqvist's dynamical treatment.
Nevertheless, it serves as an alternative hypothesis, which is illustrated in
Fig.~8.

Consider all the scalar states of Fig.~2 and assume that the $f_0$'s at
1510, 1525 and 1590 MeV are really all the same, then Tornqvist's picture
relates the nine lightest hadron states to the
$S_{q{\overline q}}=1$, $L_{q{\overline q}}=1$ quark multiplet.
As seen in Fig.~8,
this leaves the $f_0(1370)$, the $a_0(1430)$ and the $f_0(1510)$ without a
home.
These are then extra states --- one of which may well be a glueball.
In contrast, the Amsler-Close scheme$~^{10}$, with its glueball
{\it seed} built in,
has slots for ten states. One of these is a predominantly $s{\overline s}$
scalar above 1700. This they suggest is the $f_0(1710)$ ---
the erstwhile $\theta$. Again this is one of
Achasov's  {\it metamorphoses}$~^{22}$ where the
$I=0$ state at 1710 MeV seen in $J/\psi$ decays  started out as a tensor,
 then
became a scalar and is perhaps now a mixture of the two spins$~^{13}$.
 Clarification is needed.
Even given this, the Amsler-Close picture leaves the
scalars near $K{\overline K}$ threshold,
the $a_0(980)$ and $f_0(980)$, out in the cold (Fig.~8),
together with the broad $f_0(1300)$, if this is really distinct
from the $f_0(1370)$.
One has to appeal to other {\it seedings} to account for the $a_0,\; f_0(980)$,
like $K{\overline K}$-molecules or Gribov minions.
 To me this looks unlikely.
We need a more sophisticated approach to the scalars,
as Tornqvist has highlighted,
before we can reach any more definite conclusion than there are
more scalars than can fit into a simple quark model scheme.
The $f_0(1510)$ with its width of
$\sim 120$ MeV is certainly distinct from the very broad
$K^*_0(1430)$ and $f_0(1300/1370)$ states. It is
therefore very definitely a candidate for an  {\it extra} or gluonic  scalar.
Once again information on its $K{\overline K}$ couplings will be vital
to ascertain its nature.
\vskip 5mm
\noindent {\bf 4. ?`~Tensor glueball~?}
\vskip 2mm

\noindent As soon as glueballs are mentioned, one must, of course,
 report on lattice
calculations.
These give the following numbers for  scalar and tensor glueball
masses$~^{25,26}$ in quenched QCD~:
\begin{eqnarray}
0^{++}\qquad  & & 1740 \pm 70 \;{\rm MeV}\qquad \quad \quad \ \ {\rm
IBM}\nonumber \\
\, & &1550 \pm 50 \;{\rm MeV}\qquad\qquad {\rm UKQCD}\nonumber \\
& \qquad \nonumber \\
2^{++}\qquad & &\,2270 \pm 100 \,{\rm MeV}\qquad\quad \ {\rm UKQCD}\nonumber
 \end{eqnarray}

\noindent To the outsider the different groups do not appear
to have inconsistent
predictions for the gluonic scalar, but each group argues for its value
furiously.  So much so that one has the impression  that
experimenters should be having many  sleepless nights if they do not
find results in total agreement with one lattice group or the other.
However, it is important to
reiterate that these calculations are for {\it quenched} QCD,
so that there is no coupling of the primitive glueball
to quarks.  We have seen from Tornqvist's calculation that the coupling
of scalars to two pseudoscalars may have a dramatic effect on the
behaviour of the scalar propagator and though this was in the quark sector,
it  may nevertheless have some significance for the glueball case too.
Nonetheless, a scalar at 1510 MeV seems to me  quite consistent with lattice
estimates.

Work on the lattice also leads us to  expect a tensor glueball above 2 GeV
and at this meeting we heard Jin$~^{27}$ present evidence from
 BES of a candidate
the $\xi(2230)$. This state was previously seen  as a narrow spike
in $J/\psi\to \gamma K^+K^-$ and $K_sK_s$ by Mark III$~^{28}$.
 What BES have added is not
really a more pronounced signal, but rather a consistent structure in
more radiative $J/\psi$ decay channels~: in $\pi^+\pi^-$, $K^+K^-$,
$K_sK_s$ and ${\overline p}p$$~^{29,27}$. All are consistent with a resonance
of mass $2235 \pm 10$ MeV and width of $20 \pm 12$ MeV. Now it is argued that
this is a tensor glueball candidate$~^{27}$. However, it is quite unclear
whether it is a tensor and whether it is a glueball. $2^{++}$ is merely
assumed.
There is no doubt the state is narrow, but what about its branching ratios~?
 From the lack of signal for the $\xi$ in LEAR's PS185 experiment$~^{30}$,
 one can infer that
$${\rm Br}(\xi\to {\overline p}p)
\cdot {\rm Br}(\xi\to K{\overline K})\,<\, 10^{-4}
\quad .$$
Combining this with  their own results, BES$~^{27}$ infer
$${\rm Br}(\xi\to\pi^+\pi^-)\;,\;{\rm Br}(\xi\to K^+K^-) \,<\, 2\%\quad .$$
These are very small, but in fact they are very much in keeping with
the branching ratios for $\chi_{c0}$ and $\chi_{c2}$, whose decays are believed
to proceed through intermediate gluons. Before any conclusions
can be drawn, one must determine what the other $\sim 90\%$ of decays are.
Are any of these dominant~? There are those that have argued that a glueball
should have important $\eta\eta$, $\eta\eta'$ and $\eta'\eta'$ decays$~^{31}$~?
Others have suggested that a glueball being a flavour-singlet
would have branching ratios in well defined fractions, so that
$K{\overline K}\,:\,\pi\pi\,:\,\eta\eta\,:\,\eta\eta'\;
\simeq\;4\,:\,3\,:\,1\,:\,0$. Even its quantum numbers are uncertain.
Indeed, the $\xi(2230)$ sits exactly at
$\Lambda{\overline \Lambda}$ threshold ---
what is its relation, if any, with that channel~?
All this we need to know before we can conclude that the $\xi(2230)$
is a tensor and is a glueball.
{\it Prima facie} evidence that it is not a $q{\overline q}$ state is
provided by its very narrow width.
A simple OZI suppression rule would suggest that the width
of a tensor glueball
is related to that of the $\chi_{c2}$ and a typical
$q{\overline q}$ tensor, like
the $f_2(1270)$, by
$$ \Gamma(G)\,\simeq\,\left[ \Gamma(\chi_{c2})\,
 \Gamma(f_2) \right]^{1/2}\quad .$$
In round numbers the width of the $\chi_{c2}$ is 2 MeV, the $f_2$ is 200 MeV,
so one would expect $\Gamma(G) \simeq 20$ MeV, with which BES, of course,
agree.
All this is most intriguing, but we  clearly
 need much more  experimental information.

Thus, we have had two glueball candidates presented at this conference~:
the scalar $f_0(1510)$ and a tentatively tensor $\xi(2230)$.
 We see that they have quite different
characteristics.  The scalar has a typical
hadronic width of 120 MeV or so, while
the $\xi$ is very narrow $\sim 20$ MeV.
Why should these states be so different, if
they are both glueballs~?  The naive argument is to look at their position
relative to the lowest $q{\overline q}$ nonet.
Whether the scalar nonet, Fig.~8,
is as Tornqvist$~^{9}$ has it or is that of
Amsler and Close$~^{10}$, the $f_0(1510)$ is
very nearby and naturally mixes with these states.  Consequently, a
gluonic state would mix with  the $q{\overline q}$ states
and thereby decay readily into hadrons with a typical 100 MeV width.
In contrast, the $\xi(2230)$ is well above the well-known tensor nonet
and the mixings are consequently much smaller and a purer gluonic hadron
with a narrow width may result.  Of course, this has to be backed by realistic
dynamical calculations.  It begs one immediate question, which is :
yes, a tensor glueball at 2.2 GeV may mix only a little with  the ground state
tensors, but why could it not mix more with nearby radially excited states and
so become broader too~? We know so little about radially excited states, that
it is difficult to answer this sensibly.

Tests of the nature of the $f_0(1510)$ and of the $\xi(2230)$ are essential.
Information on their couplings in $J/\psi$ radiative decays
are the basis of the
\c Cakir/Farrar test$~^{32}$ and their widths into $\gamma\gamma$ would allow
the Chanowitz {\it stickiness} test$~^{33}$ to be applied ---
these are for the future.
\vskip 5mm
\noindent {\bf 5. Decays}
\vskip 2mm
\baselineskip=7.2mm

\noindent Now let us turn to the topic of decay systematics. Ackleh,
Barnes and Swanson$~^{34}$
have persuasively shown how hadron decays by the creation of a $q{\overline q}$
pair in a $^3 P_0$ state beautifully correlates a large amount of
information.  Thus it predicts the right $S/D$ ratio for $b_1 \to\omega\pi$
and provides a very interesting rule that a state with $q{\overline q}$
spin of zero cannot decay into two similar quark spin zero mesons, i.e
$S_{q{\overline q}}=0\,\not\to\,S_{q{\overline q}}=0\,+\,S_{q{\overline q}}=0$.
This very nicely explains why the
\begin{eqnarray}
\pi_2(1670)\,\not\to&\, b_1 \pi\qquad\quad ,\nonumber \\
\;\to&\, \rho \pi\;,\; f_2\pi\quad ,\nonumber
\end{eqnarray}
which hold experimentally.
The theoretical basis for this picture is that pair creation
by the scalar confining part of  the
inter-quark potential, Fig.~1, produces the $q{\overline q}$  in a $^3 P_0$
state.   Indeed, the relative magnitudes of all decays are
 predicted. All this works remarkably well
  --- except for the scalar sector, perhaps not surprisingly..

 To deal with this, Ackleh et al.$~^{34}$
 add to the $^3P_0$ component a contribution from  just one gluon exchange
 even though the  coupling has to be large (cf. Fig.1 at large distances).
 They find
 that  this simple addition eases the problem in the scalar decays, while
leaving everything else unchanged. Of course, this cannot be the complete
story.
 The whole problem of
soft physics is non-perturbative and so the appropriate framework
for solving these issues is through the study of Bethe-Salpeter amplitudes,
which automatically include the non-perturbative behaviour
 of quark and gluon propagators
and vertices that satisfy the Schwinger-Dyson equations.
I want to do nothing more than
advertise the recent progress in this approach$~^{35}$.
 One of its successes is its
ability to incorporate the Goldstone nature of the pion in a natural way.
Chiral symmetry breaking is an important feature of the
real world$~^{36}$,
which is far from obvious in a simple $q{\overline q}$ picture of the pion.
\vskip 5mm
\noindent {\bf 6. The $S$-matrix determines Physics.
Does Physics fix the $S$-matrix~?}
\vskip 2mm

\noindent I now want to comment briefly on the key problem of
getting physics  out of experiments.
It is clear that physics predicts uniquely what experiment sees.
However, given experimental information the extraction of
physics from this is fraught with
ambiguity.  This is one of the reasons why it is essential
to have many sources
of information focussing on the same physics issues.
Thus, we want to use all
of $e^+e^-$ annihilation, peripheral $\pi N$ and $K N$ scattering,
central production in $\pi p$ and $pp$ collisions,
 ${\overline p}p$ annihilation
and $J/\psi$ decays to learn about the hadron spectrum. At this conference
the results from LEAR have rightly played a central role, but they on their own
are not enough.

The standard way to analyse ${\overline p}p$ annihilation into
some multi-hadron final state, e.g. ${\overline p}p\to ABC$, is to use the
Isobar model.  This assumes that resonances only occur in the two-body channels
$AB$ with $C$ as a spectator, or $BC$ with $A$ as a spectator or $CA$ with
$B$ as a spectator. These two body channels are often $\pi\pi$, for example.
Unitarity  requires that the coupling of any resonance to $\pi\pi$,
however produced, must be universal, Fig.~3 again.
Then the ${\overline p}p$ amplitude, $\cal F$,
is intimately related to the amplitude, $\cal T$,
for $\pi\pi\to\pi\pi$ with the same quantum numbers.  This means
for the {\it cogniscente} that
$P$, or $Q-$ vectors$~^{37,38}$
( or coupling functions $\alpha$ $~^{39}$) which relate
$\cal F$ to $\cal T$ must be real, since by
the isobar assumption the third final state particle is
a spectator. Of course, the isobar model is not exact and so the relation is,
in principle, not so simple.
However, it is not obvious that just making the vectors $P$, $Q$ or $\alpha$
complex, as is often assumed,  is the only consequence.  Multi-body
final state interactions are more complicated than that.

Though the Crystal
Barrel data may be  beautifully described making such assumptions,
there are indications that the world may indeed be a more dangerous place.
Chris Pinder$~^{11}$ described the Crystal Barrel analysis of their
${\overline p}p\to\eta 3\pi$ data.  There he reported that 60\% of the events
were 4-body phase space.  Is it just an accident that this channel alone
needs multi-body interactions and that they are not present in any other~?
Maximum likelihood analysis of very many channels does
 show that only 2-body (isobar) interactions are needed,
but is that the only criterion for deciding what clues nature is offering~?
More theoretical and phenomenological work is needed. Despite these
potentially serious {\it caveats} about analyses, the beautiful data
from LEAR, and Crystal Barrel in particular, have had a dramatic
 impact on this field. It is tragic that this must end so soon.
\vskip 5mm
\newpage
\noindent {\bf 7. Conclusions}
\vskip 2mm

\noindent At {\it Hadron'95} glueball candidates have been sighted.
We need to await the next meeting before we can be sure of
all the  details, but
the $f_0(1510)$ and the $\xi(2230)$ presently
  fail to fit into $q{\overline q}$ multiplets. They are certainly
  candidates for that something extra --- the glue at the the core of QCD.
  Time will tell.
\vskip 5mm
\noindent{\bf Acknowledgements}
\vskip 2mm
\noindent It is a pleasure to thank the organizers of this conference,
 particularly
Sandy Donnachie and Judith McGovern, for their invaluable assistance.
I acknowledge
helpful discussions with Ted Barnes, Frank Close, Gaston Guttierez,
Andrew Kirk and David Morgan
and the incisive questioning of my fellow-summariser, Suh-Urk Chung,
 which made it all such good fun.
Travel
 support  was provided by the
  EC Human Capital and Mobility Programme
 network {\it EURODA$\Phi$NE} under grant CHRX-CT920026.


\vspace*{1cm} FIGURES CAN BE OBTAINED FROM THE AUTHOR

\end{document}